\pdfoutput=1
\documentclass[twocolumn,prb,showpacs,aps,superscriptaddress,reprint]{revtex4-1}

\usepackage[english]{babel}
\usepackage[ansinew]{inputenc}
\usepackage{times}
\usepackage{graphicx}
\usepackage{graphics}
\usepackage{amsmath}
\usepackage{amsfonts}
\usepackage{amssymb}
\usepackage{amsbsy}
\usepackage{dsfont}
\usepackage{epstopdf}
\usepackage{makeidx}
\usepackage{subfigure}
\usepackage{color} 
\usepackage{pgf}
\usepackage{bm}
\usepackage{tikz} \usepackage[normalem]{ulem}

\newcommand{\hcoeff}[9]{H\!\!{\tiny{\begin{array}{l}#1 #2 #3 \\ #4 #5 #6 \\ #7 #8 #9 \end{array}}}}

\makeindex
\begin{document}

\title{Extending the Accuracy of the SNAP Interatomic Potential Form}

\author{M. A. Wood}
\affiliation{Center for Computing Research, Sandia National Laboratories, Albuquerque, New Mexico 87185, USA}

\author{A. P. Thompson}
\affiliation{Center for Computing Research, Sandia National Laboratories, Albuquerque, New Mexico 87185, USA}

\date{\today}

\begin{abstract}
The Spectral Neighbor Analysis Potential (SNAP) is a classical interatomic potential that expresses the energy of each atom as a linear function of selected bispectrum components of the neighbor atoms. An extension of the SNAP form is proposed that includes quadratic terms in the bispectrum components. The extension is shown to provide a large increase in accuracy relative to the linear form, while incurring only a modest increase in computational cost. The mathematical structure of the quadratic SNAP form is similar to the embedded atom method (EAM), with the SNAP bispectrum components serving as counterparts to the two-body density functions in EAM. The effectiveness of the new form is demonstrated using an extensive set of training data for tantalum structures. Similarly to artificial neural network potentials, the quadratic SNAP form requires substantially more training data in order to prevent overfitting. The quality of this new potential form is measured through a robust cross-validation analysis.
\end{abstract}


\maketitle
\section{\label{intro}Introduction}

Simulation methods such as classical molecular dynamics or Monte Carlo sampling are powerful tools for examining the behaviors of materials that originate on the atomic scale.  
Using scalable algorithms and large parallel computers it is now possible to routinely simulate the structural evolution of chunks of material containing millions of atoms on timescales approaching one microsecond.  
This is possible because, in many situations, an explicit quantum mechanical treatment of the electronic degrees of freedom can be avoided, by defining a classical interatomic potential function (IAP) that depends only on the atomic degrees of freedom.  
Typical IAP forms provide short-ranged explicit expressions for the forces on the atoms that can be efficiently decomposed into independent parallel computations. 
As a result, MD simulations are many orders of magnitude faster than the corresponding quantum mechanical simulation for systems composed of tens or hundreds of atoms.  
Moreover, while the cost of even a single force evaluation becomes prohibitively expensive for quantum calculations involving more than a few hundred atoms, the computational speed of classical MD simulations remains largely independent of the total number of atoms, provided that large problems are executed on an equally large number of processors.\cite{Griebel2007} 

The extent to which MD simulations provide an accurate description of real materials is largely dependent on how well the IAP approximates the true quantum mechanical energy of the system.\cite{Finnis2003}
An IAP is essentially a reduced order model that takes the full quantum many-body problem of ionic and electronic interactions and casts it into computationally tractable expressions for the total energy as the sum of individual atom contributions.\cite{Brenner2000}
However, this simplification can be done in many different ways, giving rise to the numerous different types of IAP that are seen in the literature. 
The simplest, and computationally most efficient express the energies and forces solely in terms of pair-wise interactions between nearby atoms.\cite{LJ1931, Rappe1992} 
In recent years, there has been a rise in the use of more complex and costly potentials.\cite{Plimpton2012}
These advancements have enabled MD simulations of chemical reactions in organic systems\cite{vanduin2001,  Tersoff1988, Brenner2002} and electric charge polarization in insulating materials.\cite{Rappe1991, Shan2010}

Nearly all of these IAP have mathematical forms that are physically and chemically motivated. They exploit our understanding of the nature of a chemical bond or mechanical deformation of solids to formulate the interatomic potential. 
For example, the embedded atom method (EAM) decomposes the potential energy of each atom into pair interactions with neighbors and a many-body embedding energy that depends on the local mean electron density due to neighbor atoms.\cite{Daw1983, Daw1984}
Extensions of EAM use higher order moments of the local electron density to represent polarization and directional bonding effects.\cite{Baskes1992}
Bond order potentials like Tersoff\cite{Tersoff1988} and ReaxFF\cite{vanduin2001, Senftle2016} simplify covalent bonding into smooth functions of interatomic distance and atomic coordination.
The finer details of the chemical environment around a pair of bonded atoms can be described by more complex energy expressions that describe specific bonding patterns.\cite{Stuart2000, Brenner2002} 

An alternative to this approach relies not on advances in physics and chemistry theory, but rather data science and artificial intelligence.
Instead of developing theory-inspired functions for IAPs, the approach of machine-learned interatomic potentials (ML-IAP) adopts a very general mathematical representation of the potential energy that satisfies just a few essential physical requirements. These IAP contain many free parameters that are trained to reproduce features of the high fidelity electronic structure calculations.\cite{Behler2007,Behler2016}
Because of this generality, it is thought that ML-IAP are more flexible and are capable of accurately describing a wide range of chemical systems, provided that sufficient reference data is used to train and test. 

The improved accuracy of these ML-IAP comes at the price of increased computational cost.  This cost is incurred both in the process of generating the training data for the ML-IAP and also when it is used to run large-scale MD simulations.  Nonetheless, because generation of the training data requires large numbers of quantum electronic structure calculations with small numbers of atoms, while the ML-IAP force calculations scale even better than conventional IAPs, the increase in computational cost is greatly mitigated by the increasing abundance of low-cost massively-parallel computational resources. 

In conceiving and refining our particular approach to ML-IAP, we have sought to find a good balance between accuracy and computational cost. These considerations lead us to construct the Spectral Neighbor Analysis Potential (SNAP) model.  The SNAP method was shown to provide a good description of a range of different properties of various solid phases of tantalum, and also the liquid structure.\cite{Thompson2015}  In this paper, we demonstrate how several fairly simple extensions to the SNAP model significantly improve both training and cross-validation errors with only modest increase in computational cost. 

\section{\label{model}Regression Methodology}
The purpose of any ML-IAP is to express the potential energy surface (PES) in terms of a set of scalar functions known as descriptors.  
The descriptors are  somewhat arbitrary functions that depend on the local atomic environment of an atom.  
Each atomic environment, expressed as Cartesian coordinates of the neighbor atoms relative to the central atom, is first mapped to a point in the descriptor space. 
That point in turn is mapped to a value of energy.  
The overall effectiveness of the ML-IAP is sensitive to the choice of descriptor space.
As a minimal requirement, the descriptor space should be invariant under rotation, permutation, and translation of the atomic environment.\cite{Behler2007, Ferre2015}
In addition, for multi-element potentials, the descriptors should distinguish between different atom types.\cite{Kobayashi2017, Takahashi2017}
In recent years, many different types of descriptors have been proposed.\cite{Bartok2013a, Rupp2012, Montavon2013} 
More complicated descriptor spaces include basis expansions based on symmetries in the local environment around each atom.\cite{Szlachta2014, Lilienfeld2015, Podryabinkin2017, Bartok2015} 

The bispectrum components, which are the focus of this work, fall into this category.\cite{Thompson2015, Bartok2010, BartokThesis}. They have also been used as descriptors in several recent studies by other groups.\cite{Chen2017, Shweta2018}
The question of which approach to describing atomic neighborhoods is most effective is an active area of research. For example, there is some evidence that direct pairwise comparison of atomic neighborhoods using a similarity measure is more effective than trying to resolve the atomic neighborhood using invariants extracted from a basis function expansion, such as the bispectrum components.\cite{Bartok2015}   However, basis expansion invariants have the advantage of being directly amenable to linear regression methods.
 
ML-IAP must be trained using reference data in order to construct a model form that connects the descriptor space to the target feature space i.e. known energies and forces from a higher fidelity model.\cite{Csanyi2004, Zhenwei2015, Deringer2017}
This training step can be approached in a variety of ways.
One powerful class of method is the artificial neural network (ANN).\cite{Bishop1995, Behler2011, Behler2014}  
This uses multiple convolutions (layers) of the descriptor space to capture the most important features. The weight assigned to connections between nodes in successive layers are free variables in the fit. 
While shown to produce some of the most accurate representations of training data,\cite{Lubbers2017, Smith2017} a typical ANN requires significantly more training data than other model forms, and contains $\mathcal{O}(10^{3}-10^{5})$ fit parameters, which can make it difficult to train.
Other model forms interpolate between data in the training set through some similarity metric for unknown structures.\cite{De2016, Barker2017,Szlachta2014} However, these ML-IAP are computationally expensive to use for MD simulations, because the cost scales with the size of the training set.\cite{Ischtwan1994, Crespos2003, Crespos2004}

Lastly there exists a set of linear regression models that reduce the dimensionality of the training set size to that of the descriptor space via a linear transformation.\cite{Bishop2007} 
These linear models are a special case of ANN with only a single convolution layer, i.e. the number of fitting terms in the regression is equal to the size of the descriptor space.
Linear models can also be viewed as a special form of Gaussian process regression in which a dot product kernel is used for the covariance matrix. \cite{Rasmussen2006} 

The mapping from the descriptor space ($D$) to the training space ($T$) is determined by the vector of linear coefficients $\boldsymbol\beta$, chosen to minimize the Euclidean norm distance between training and predicted points in the training space.
\begin{equation}
{\min}(||{\bf w}\cdot D \boldsymbol\beta- T||^{2}-\gamma_{n}~||\boldsymbol\beta||^{n})
\label{euclid}
\end{equation}
A regularization penalty (second term in Eq. \ref{euclid}) of order $n$ can be added to the linear least squares (LLS) term. Solutions with $n=1$ regularization enforce sparsity of the $\boldsymbol\beta$ solution, known as LASSO solutions.\cite{Tibshirani1996}
This can be used to isolate key features of the descriptor space, but is rarely used to construct ML-IAP because the training space is often much larger than the descriptor space. 
Similarly, solutions with $n=2$ will penalize large values of $\boldsymbol\beta$ which are hallmarks of an overfit solution, known as Tikhonov regularization.\cite{Tikhonov2013}
In this work an overfit solutions is the result of a lack of diversity in the training set relative to the number of free variables used in the fit.
Avoiding overfit solutions that are also capable of accurate out-of-sample prediction is critical for producing a viable ML-IAP for use in MD. 
A simple means to improve the fit quality is to add more degrees of freedom, but if this is done haphazardly the overall transferability of the potential will be sacrificed. 
Additional weights can be applied to specific terms of the descriptor space through the vector ${\bf w}$.

\section{\label{tradeoff}Linear SNAP}
In this section, we discuss some general practical concerns around the use of for ML-IAP.
While there has been significant evidence that ML-IAP are capable of producing high accuracy results with respect to DFT data, a balance 
must be struck between the effort required to generate training data and the computational cost of running a simulation.
To many end users, limitations on computing resources are an important factor when selecting an IAP for MD simulation. 
In addition, for developers of IAPs, significant time is required to train and test candidate potentials.
Therefore, regardless of how elegant the model form or absolute accuracy of the method, consideration of these front- and back-end computational costs must be made. 

It is difficult to precisely quantify the time spent constructing training data because this will largely depend on the details of what the ML-IAP is intended for. 
For example, crystalline properties of metals can be calculated quickly within DFT, because perfectly periodic structures can be represented with just a few atoms.
However, extrapolation of such a potential to configurations such as defects and surfaces will result in large errors since there is nothing in this limited training set to constrain the model in those regions of the PES. 
Therefore a diverse training set is preferred in practice so that end usage of the potential is less susceptible to large extrapolation errors.
This, of course, adds to the upfront cost of parameterizing the IAP. 

The dominant cost of running an MD simulation is the evaluation of the descriptors, from which the corresponding force contributions can be obtained.
We outline here the structure of the SNAP ML-IAP in terms of the underlying descriptor space.\cite{Thompson2015}
The total potential energy of a configuration of atoms is first written as the sum of SNAP energy contributions associated with each atom, combined with a reference
potential
\begin{equation}
E({\bf r}^{N})=E_{ref}({\bf r}^{N})+\sum_{i=1}^{N}E_{\scriptscriptstyle{SNAP}}^i,
\label{snapE}
\end{equation}
where ${\bf r}^N$ is the vector of $N$ atom positions in the configuration. $E$ and $E_{ref}$ are the total and reference potential energies, respectively.
$E_{\scriptscriptstyle{SNAP}}^i$ is the SNAP potential energy associated with a particular atom $i$,
and depends only on the relative positions of its neighbor atoms.
Including a reference potential is advantageous because it can correctly represent known limiting cases of atomic interactions, leaving the SNAP contribution to capture many-body effects.
The ZBL pair potential\cite{ZBL1982} is a convenient choice, because it captures the known short-range repulsive interactions between atomic cores that are not well represented by quantum calculations. 

\begin{figure}[t]
\includegraphics{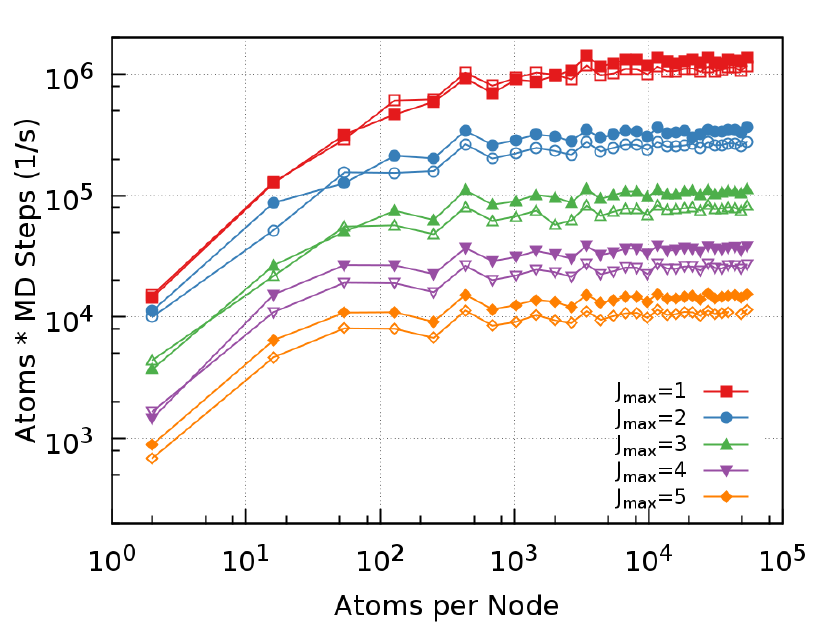} 
\caption{\label{timing} Computational speed versus atom count for linear (closed symbols) and quadratic (open symbols) SNAP potential forms with increasing values of $J_{max}$ and correspondingly larger descriptor spaces (see Table \ref{descriptors}). Timings were made using a single Intel Broadwell node. For each potential, the speed in atom$\cdot$timesteps/s becomes roughly constant for sufficiently large atom count, reflecting the linear scaling of the underlying LAMMPS\cite{Plimpton1995} neighbor list algorithm.}
\end{figure}

The construction of the SNAP component of the potential energy in terms of the bispectrum components 
follows the same approach described in
Ref. [\onlinecite{Thompson2015}], which we briefly summarize here.
The SNAP formulation begins with a very general characterization of the
neighborhood of an atom.
The density of neighbor atoms at location $\textbf{r}$ relative to a central atom $i$ can be considered as a sum of $\delta$-functions located in a three-dimensional space:
\begin{equation}
\rho_i ({\bf r}) = \delta({\bf r}) + \sum_{r_{i'} < R_{cut}}{f_c(r_{i'}) w_{i'} \delta({\bf r}-{\bf r}_{i'})}
\end{equation}
where ${\bf r}_{i'}$ is the position of neighbor atom $i'$ relative to central atom $i$.  The $w_{i'}$ coefficients are dimensionless weights that are chosen to distinguish atoms of different types, while the central atom is arbitrarily assigned a unit weight.  The sum is over all atoms $i'$ within some cutoff distance $R_{cut}$.  The switching function $f_c(r)$ ensures that the contribution of each neighbor atom goes smoothly to zero at $R_{cut}$.  
Typically, this density function is expanded in an angular basis of spherical harmonics combined with an orthonormal radial basis. \cite{Bartok2013b}
Instead, we use an idea originally proposed by Bart{\'{o}}k et al.\cite{Bartok2010}, in which the radial coordinate $r$ is mapped on to a third
angular coordinate $\theta_0 = \theta_0^{max} r / R_{cut}$.  
Each neighbor position $(r, \theta, \phi)$ is mapped to $(\theta_0, \phi, \theta)$, a point on the unit 3-sphere.  
The natural basis for functions on the 3-sphere is formed by the 4D hyperspherical harmonics $U^j_{m,m'}(\theta_0,\theta,\phi)$, defined for $j=0,\frac{1}{2},1,\ldots$ and 
$m,m' = -j,-j\!+\!1,\ldots,j\!-\!1,j$~\cite{Varshalovich1987}.  
The neighbor density function can now be expanded in the basis of hyperspherical harmonics $U^j_{m,m'}$.  
Because the neighbor density is a weighted sum of $\delta$-functions, each expansion coefficient is a sum over discrete values 
of the corresponding basis function evaluated at each neighbor position 
\begin{equation}
u^j_{m,m'} = U^j_{m,m'}(0) + \!\!\!\!\!\!\!\!\sum_{\tiny{r_{ii'} < R_{cut}}}{\!\!\!\!f_c(r_{ii'}) w_{i'} U^j_{m,m'}(\theta_0,\theta,\phi)} 
\label{eq:u}
\end{equation}
The bispectrum components are formed as the scalar triple products of the expansion coefficients
\begin{equation}
B_{j_1,j_2,j}  = \\
\!\!\!\!\sum_{m,m'} u^{j*}_{m,m'} \!\!\!\!\sum_{\tiny\begin{array}{l} \!\!m_1,m'_1 \\ m_2,m'_2 \end{array}}
\hcoeff{j}{m}{m'}{j_1}{\!m_1}{\!m'_1}{j_2}{m_2}{m'_2}
u^{j_1}_{m_1,m'_1} u^{j_2}_{m_2,m'_2}
\label{eq:bispectrum}
\end{equation}
where * indicates complex conjugation and the constants
$\hcoeff{j}{m}{m'}{j_1}{\!m_1}{\!m'_1}{j_2}{m_2}{m'_2}$
are Clebsch-Gordan coupling coefficients for the hyperspherical harmonics.
The bispectrum components are real-valued and invariant under rotation~\cite{BartokThesis}.  They are
also symmetric in the three indices $j_1, j_2, j$ up to a normalization factor.~\cite{Thompson2015}
They characterize the strength of density correlations at three points on the 3-sphere.  The lowest-order components describe the coarsest features of the density function, while higher-order components reflect finer detail.  The number of distinct bispectrum components with indices $j_1, j_2, j$ less than or equal to $J_{max}$ is given in Table~\ref{descriptors}.  By increasing the value of $J_{max}$ we can, in principle, systematically increase the accuracy of the SNAP potential, at the price of greater computational cost, as shown in Figure~\ref{timing}.  For a particular choice of $J_{max}$, we can list the $K$ bispectrum components in some arbitrary order as ${B}_{1},\ldots,{B}_{K}$.  The SNAP energy of an atom is written
as a linear function of these $K$ bispectrum components 
\begin{eqnarray}
E_{\scriptscriptstyle{SNAP}}^i & = & {\beta}_{0}+\sum_{k=1}^{K}{\beta_k}({B}_{k}^{i}-{B}_{k{0}}^{i}) \\
& = & {\beta}_{0}+\boldsymbol\beta\cdot{\bf B}^{i}
\label{eq2}
\end{eqnarray}
where ${B}_{k}^{i}$ is the $k$th bispectrum component of atom $i$ and $\beta_k$ is the associated linear coefficient, a free parameter in the SNAP model.  As a computational convenience, the contribution of each bispectrum component to the SNAP energy is shifted by the contribution of an isolated atom, $\beta_k {B}_{k0}^{i}$, so that the SNAP energy of the isolated atom is equal to $\beta_0$ by construction.
Similarly, the force on each atom $j$ due to the SNAP potential can be expressed as a weighted sum over the derivatives w.r.t. ${\bf r}_j$ of the bispectrum components of each atom $i$.
\begin{equation}
{\bf F}^{j}_{\scriptscriptstyle{SNAP}}=-\nabla_{j}\sum_{i=1}^{N}E_{\scriptscriptstyle{SNAP}}^i=-\boldsymbol\beta{\cdot}\sum_{i=1}^{N} \frac {\partial {\bf {B}}^{i}}{\partial {\bf r}_{j}}
\label{eq3}
\end{equation}
In this way, the total energy, forces, and also the stress tensor, can be written as linear functions of quantities related to the bispectrum components of the atoms. 
In addition to shifting the bispectrum components by ${B}_{k0}^{i}$, it also makes sense to set ${\beta}_{0}=0$, constraining the potential energy of an isolated atom to be zero.  This ensures that SNAP correctly reproduces the cohesive energy of the reference solid structure, an important  physical attribute of any general purpose interatomic potential.
\begin{table}[b]
\caption{Number of bispectrum components (descriptors), linear coefficients, and quadratic coefficients for increasing values of $J_{max}$.}
\begin{ruledtabular}
\begin{tabular}{p{1.9cm}p{2.0cm}p{1.9cm}p{1.9cm}}
$J_{max}$&\hspace{-2em}Descriptors&\hspace{-2em}Linear&\hspace{-2em}Quadratic\\
&$K$&\hspace{-2em}Coefficients&\hspace{-2em}Coefficients\\
\hline
1&5&5&15\\
2&14&14&105\\
3&30&30&465\\
4&55&55&1,540\\
5&91&91&4,186\\
$J$&\hspace{-3em}$\frac{(J+1)(J+\frac{3}{2})(J+2)}{3}$&$K$&$\frac{K(K+1)}{2}$\\
\end{tabular}
\end{ruledtabular}
\label{descriptors}
\end{table}

Using a linear regression model such as LLS whose solution yields $\boldsymbol\beta$, the cost of SNAP is decoupled from the size of the training set, and will only depend on the number of bispectrum components calculated for each atom. 
Table \ref{descriptors} shows how the size of the descriptor space $K$ grows as a function of $J_{max}$, the maximum order of the bispectrum components.
It is important to note the ratio of fitting coefficients to the size of the descriptor space for either SNAP form.
The quadratic SNAP will be be able to capture more features of the training set due to the increased number of free fitting variables.
Since the SNAP potential energy is a linear function of these $K$ descriptors, the number of coefficients that are solved for in the linear regression ($\boldsymbol\beta$ in Eq. \ref{euclid}) is equal to the size of the descriptor space. 

The computational cost of SNAP depends strongly on the size of the descriptor space.\cite{Trott2014} Figure \ref{timing} captures this behavior for values of $J_{max}$ ranging from 1 to 5. 
We measure computational cost by running a short MD simulation using LAMMPS\cite{Plimpton1995,lammpsweb} for each SNAP potential on a single Intel Broadwell node of Sandia's Serrano computing cluster. The node consists of two 2.1 GHz Intel Broadwell E5-2695v4 processors with 18 physical cores each.
At runtime, the performance of each SNAP potential was optimized by varying the number of MPI tasks while adjusting the number of OpenMP threads per MPI task to keep the total number of threads on the node equal to 36, the total number of cores. 
We did not enable hyperthreading, non-standard NUMA settings, or allow some cores to remain idle.  
We repeated this over a wide range of atom counts, because the optimal number of atoms per node is different for different SNAP potentials.\cite{lammpssnap}
The closed symbols in Figure \ref{timing} correspond to the linear form of SNAP while open symbols refer to the quadratic formulation of the SNAP energy function which will be discussed further in the next section. 
Clearly, as the number of bispectrum components increases, the speed of the MD calculation diminishes.
In going from $J_{max}=1$ to $J_{max}=5$, the size of the descriptor space increases by about a factor of 20, and the computational cost increases by approximately two orders of magnitude.
For comparison, a MEAM potential\cite{Scheiber2016,Wood2017} for tungsten reached a peak speed of $5\cdot10^{6}$ atom$\cdot$timesteps/s on the same hardware, while the GAP potential for tungsten\cite{Szlachta2014, QUIP} reached a peak performance near $5\cdot10^{2}$ atom$\cdot$timesteps/s.
It is worth mentioning that the more costly GAP IAP goes beyond linear accuracy in the descriptor space, this point is expanded upon in the following section.

\section{\label{qsnap}Quadratic SNAP}
One of the potential limitations of the SNAP approach is the restriction to a linear relationship between the atom potential energy and the bispectrum components of the atom neighbor density.  
It is shown below that this limits the bodiedness or maximum complexity of energy terms to four-body. 
Just as the embedded atom method has proven to be an effective way to add many-body interactions to pair potentials\cite{Daw1983, Daw1984}, we can extend the bodiedness of SNAP potentials by adding an embedding energy term
\begin{eqnarray}
E_{\scriptscriptstyle{SNAP}}^i({\bf r}^{N}) & = & \boldsymbol\beta\cdot{\bf B}^{i} + F(\rho_i),
\label{eq4}
\end{eqnarray}
where $F(\rho_i)$ represents the energy of embedding atom $i$ into the electron density contributed by its neighboring atoms. 
Just as $\rho_i$ is written as a sum of pair contributions in EAM, we write it here as a linear function of the bispectrum components.
\begin{eqnarray}
\rho_i & = & {\bf a}\cdot{\bf B}^{i}
\label{eq5}
\end{eqnarray}
We express the embedding energy as a Taylor expansion about some reference structure with density $\rho_0$,
\begin{eqnarray}
F(\rho) & = & F_0 + (\rho-\rho_0) F^{\prime} + \frac{1}{2}(\rho-\rho_0)^2 F^{\prime\prime} + ...
\label{eq6}
\end{eqnarray}
where $F_0$, $F^{\prime}$, and $F^{\prime\prime}$ are constant values equal to the embedding function and its first and second derivatives at the reference density $\rho_0$, respectively.  
The constant and linear contributions to the embedding energy can be written as equivalent contributions to the linear SNAP energy. 
This is directly analogous to how the linear contributions to the EAM embedding energy can be written in terms of an equivalent pair potential. \cite{Daw1984}
Substituting the quadratic term into Eq. \ref{eq4} we find
\begin{eqnarray}
E_{\scriptscriptstyle{SNAP}}^i({\bf r}^{N}) & = & \boldsymbol\beta\cdot{\bf B}^{i} + \frac{1}{2} F^{\prime\prime} ({\bf a}\cdot{\bf B}^{i})^2 \\
                                               & = & \boldsymbol\beta\cdot{\bf B}^{i} + \frac{1}{2} ({\bf B}^{i})^T\cdot\boldsymbol{\alpha}\cdot{\bf B}^{i},
\label{eq7}
\end{eqnarray}
where $\boldsymbol{\alpha} = F^{\prime\prime} {\bf a} \otimes {\bf a}$ is a symmetric $K \times K$ matrix.  Eq. \ref{eq7} is simply the extension of the linear SNAP form to include all distinct pairwise products of bispectrum components and we refer to it as the quadratic SNAP model.  

In comparing different interatomic potential forms, it is instructive to consider their "bodiedness" or maximum complexity of the energy expressions generated.  Simple interatomic potentials are often expressed as truncated cluster expansions, in which the total energy is decomposed into sums over terms depending on scalar distance between pairs of atoms (two-body), angle formed by triplets of atoms (three-body), dihedral angle formed by four atoms, and so on.  This idea can be extended to more complex potential by first fully expanding the total potential energy expression into a sum of individual terms and then considering the maximum number of atom positions appearing in a single term.  Performing this exercise for the linear SNAP form requires fully expanding the sums in Eqs.~\ref{eq:u} and \ref{eq:bispectrum}.
For the SNAP energy of a particular atom $i$ we obtain individual terms of the form 
\begin{equation*}
U^{j_1*}_{m_1,m'_1}({\bf r}_{i_1}-{\bf r}_{i}) U^{j_2}_{m_2,m'_2}({\bf r}_{i_2}-{\bf r}_{i}) U^{j_3}_{m_3,m'_3}({\bf r}_{i_3}-{\bf r}_{i}) 
\end{equation*}
where ${\bf r}_{i}$, ${\bf r}_{i_1}$, ${\bf r}_{i_2}$, and ${\bf r}_{i_3}$ are the positions of atom $i$ and three neighbors $i_1$, $i_2$, and $i_3$, respectively.  The associated prefactors of $f_c(r)$ and $w_i$ have been omitted, and the mapping from ${\bf r}$ to $(\theta_0, \phi, \theta)$ is not shown explicitly.  
This shows that the maximum number of 
distinct atom positions appearing in a single term is four.  In the case of quadratic SNAP, the individual terms are pairwise products of these terms.  
Because both factors involve atom $i$, the maximum number of distinct atom positions appearing is seven 
(one central atom plus two sets of three distinct neighbor atoms).
While the maximum complexity of the energy terms generated by linear SNAP is four-body, quadratic SNAP generates terms up to seven-body.  
It is unclear to what extent limiting the bodiedness of an interatomic potential fundamentally restricts its predictive power, but the results presented below
suggests that there is a practical benefit to going being four-body terms.
The total energy remains a linear function of  $\boldsymbol{\beta}$ and $\boldsymbol{\alpha}$ which means it can still be solved using linear regression methods as in Eq. \ref{euclid}.
\begin{figure}[t]
\includegraphics{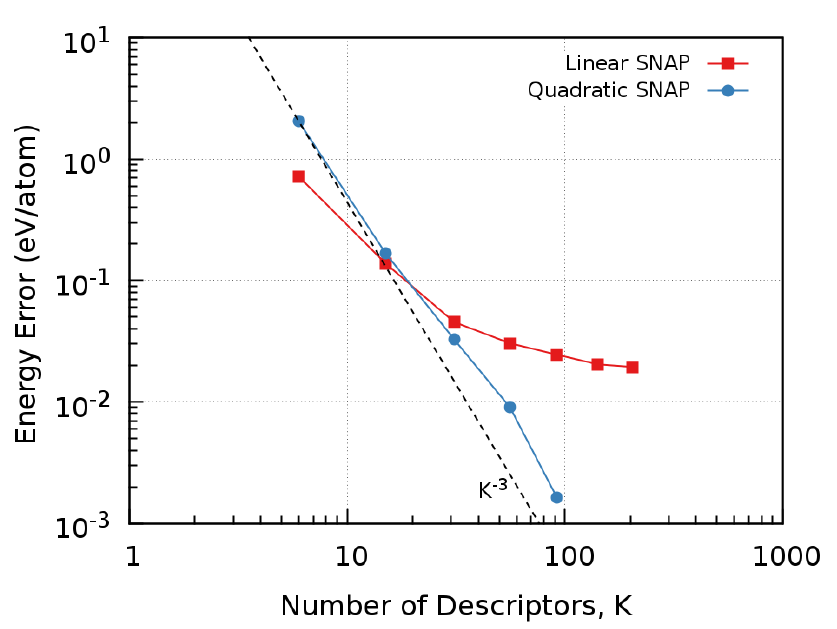} 
\caption{\label{eerr} Mean absolute energy error for SNAP models relative to the DFT training data for tantalum versus the number of descriptors $K$. The linear SNAP form (red squares) shows negligible improvements in accuracy beyond $K=30$. In contrast, the quadratic SNAP form (blue squares) exhibits continuously improving accuracy, with error decreasing in proportion to $\sim K^{-3}$ (dashed line).}
\end{figure}

The quadratic SNAP model can be viewed as a special case of an ANN whose inputs are the coordinates of neighbor atoms relative to each central atom.  
There are two hidden layers whose activation functions are the bispectrum components of each atom and the product of pairs of bispectrum components, respectively.  
The weights on the connections between layers depend on the values of $\boldsymbol{\alpha}$ and $\boldsymbol{\beta}$.  
If $\boldsymbol{\alpha}$ is set to zero, then the two layers collapse into one and the linear SNAP form is recovered.
Whereas a typical ANN uses two-body and three-body descriptors as inputs\cite{Behler2014}, the four-body SNAP descriptors require fewer hidden layers to achieve good predictive accuracy.
A clear benefit of quadratic SNAP over linear SNAP, similar to that general purpose ANN, is the introduction of nonlinear terms and a 
corresponding increase in the number of fitting coefficients, without expanding the size of the descriptor space. 

A demonstration of the accuracy gained from adding the quadratic SNAP terms is given in Figure \ref{eerr}.
For both the linear (red squares) and quadratic (blue circles) SNAP forms, the number of unique descriptors is the same for a given $J_{max}$, 
but there are significantly more coefficients to fit in the latter (see Table \ref{descriptors}).
Figure \ref{eerr} shows diminishing improvements in the accuracy of the linear SNAP form as the size of the descriptor space is increased. 
In contrast, the quadratic SNAP forms shows continued systematic improvement in accuracy over the same range\cite{sidecomment1}, with the energy error decreasing roughly in proportion to $\sim K^{-3}$.
At $J_{max}=5$ ($K=91$) there is an order of magnitude better energy error for quadratic SNAP over the linear form.
As shown in Figure~\ref{timing}, this large improvement in accuracy between model forms is  accompanied by less than a factor of two decrease in speed (atom$\cdot$timesteps/s).
In the subsequent sections we will provide more details on how SNAP potentials are optimized, as well as a discussion of how to avoid pitfalls related to overfitting and extrapolation beyond the range of training data.

Before closing this section it is worth mentioning that we see a natural extension of SNAP as a many-body correction to existing potentials.
This could be anything from a simple Lennard-Jones\cite{LJ1931} potential to a complex many-body potential such as MEAM\cite{Baskes1992} or COMB.\cite{Shan2010}
Existing empirical potentials could include an additional SNAP many-body correction, given some specific set of electronic structure data relevant to the current application.
For example, an EAM potential for a metallic system could include a SNAP correction in order to capture the relative stability of metastable crystal phases that were previously poorly represented. 
This could also be done for semi-empirical methods such as tight-binding\cite{Cawkwell2012}, where many-body repulsions are empirically added to the total energy functional. 

\section{\label{optimize}Optimization Methodology}
The need for a new IAP is driven by the scientific needs of a particular application. This means there exists a set of target properties for which the potential should be optimized. 
These quality metrics are material properties such as elastic constants, defect formation energies, melting temperature, and relative phase stability.
We consider these externally imposed quality metrics as probes of how well the generated potential describes physically important configurations that may not be well-represented in the training data. In addition, a ML-IAP must also be optimized to predict energies and forces contained in the training data.  A combination of these training and prediction error metrics constitute a set of objective functions that are used to optimize the SNAP potential.
In this section we describe in some detail the steps we take to optimize potentials using both categories of objective functions.
While these steps are discussed, a direct comparison of the accuracy between linear and quadratic SNAP will be made for elemental tantalum.\cite{Thompson2015}
\begin{figure}[t]
\includegraphics{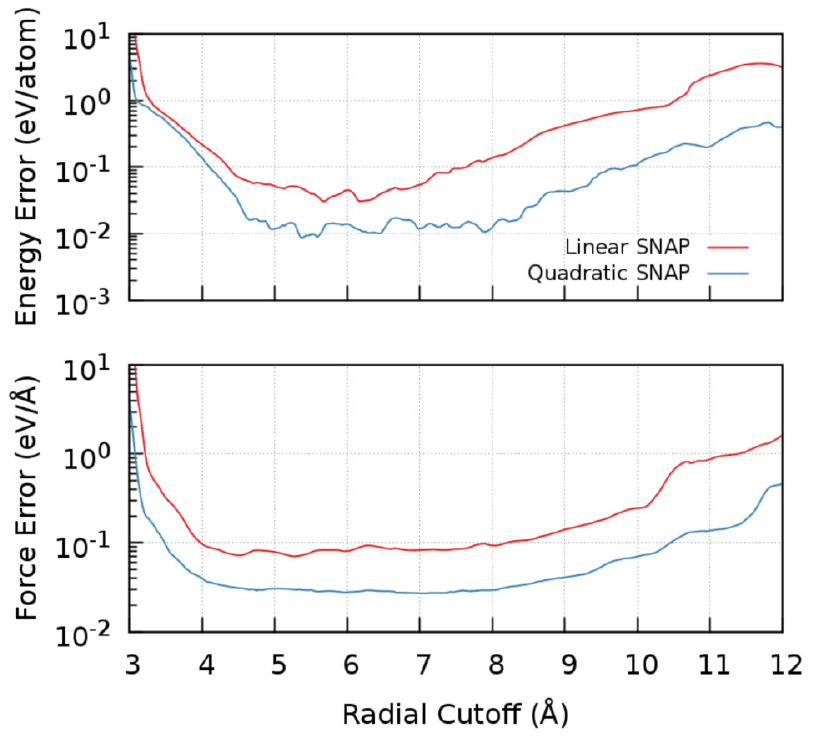} 
\caption{\label{hyper}  Training errors for the linear (red) and quadratic (blue) SNAP forms for different values of the radial cutoff distance $r_{cut}$.  {\bf Top} Mean absolute error for energy data {\bf Bottom} Mean absolute error for force data. For both energy and force errors, the quadratic form of SNAP yields lower training errors than the linear form, while the size of the descriptor space is the same in both cases ($J_{max}=4$).}
\end{figure}

\begin{figure*}[t]
\includegraphics{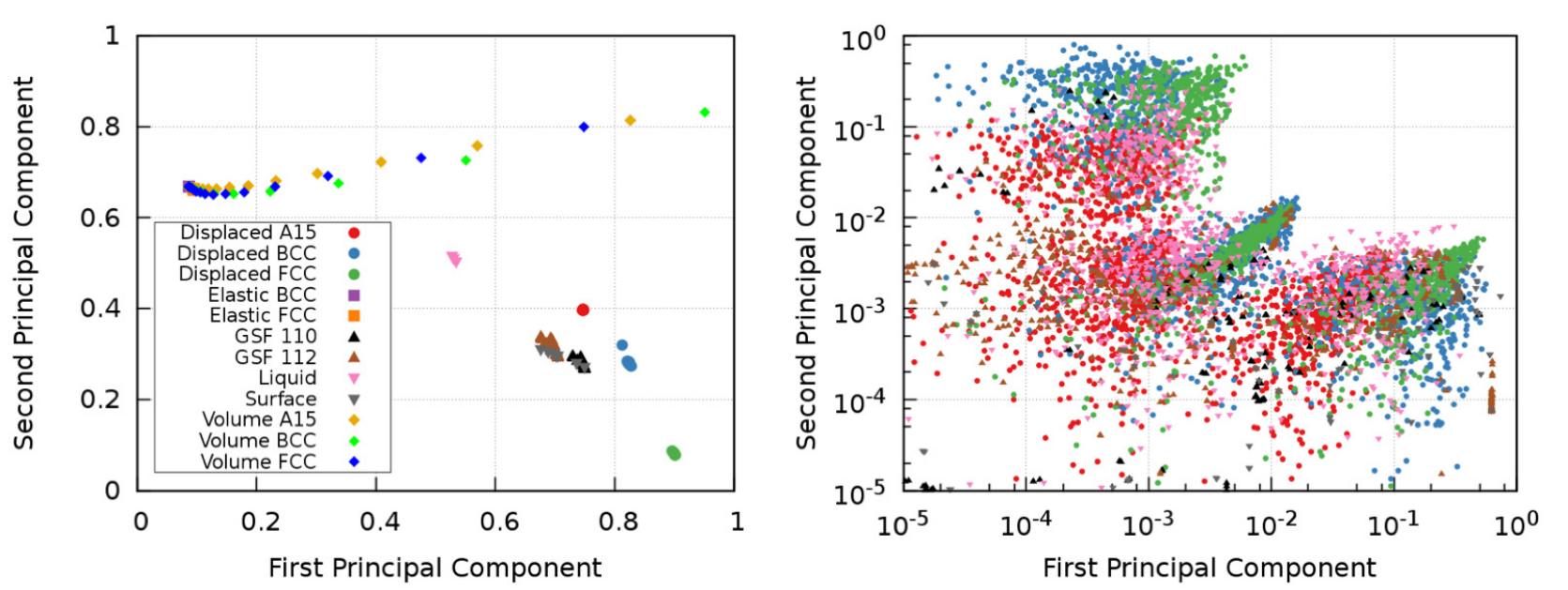} 
\caption{\label{pca} 
{\bf Left} Projection of the descriptor space for energy training data onto the first two principal components. {\bf Right} Projection of the descriptor space for force training data onto the first two principal components. }
\end{figure*}
As was mentioned in Section \ref{tradeoff}, a significant amount of expert opinion is needed in the generation of useful training data. 
In a computationally tractable way, DFT (or any high fidelity model) data must be generated that samples the most important regions of a material's PES. 
For the results presented here, we have used the previously generated training set for tantalum described in Ref. [\onlinecite{Thompson2015}].
This training set was constructed in order to train a SNAP potential to the energies and forces from DFT of both crystalline and disordered states.
A full list of the different groups of training configurations is given in the legend of Figure \ref{pca}.
The \emph{Displaced A15, BCC and FCC} groups correspond to randomly displaced atoms of a large (approx. 50 atoms) super cell of each crystal phase, these groups contribute 27 energy and 4482 force training points. 
Groups \emph{Elastic BCC and FCC} are unit cells of either phase with the simulation cell perturbed from the equilibrium structure, these contribute 200 energies and 1200 forces in total. 
Similarly, \emph{Volume A15, BCC and FCC} are simply volumetric compressions and expansions of the equilibrium crystal structure, 81 energy and 1218 force training points are attributed to these groups. 
The \emph{Liquid} group contributes 3 energy and 4203 force points, \emph{Surface} group adds another 7 energy and 630 forces.
Lastly, the \emph{GSF 110 and 112} are generalized stacking fault calculations on the $(110)$ and $(112)$ crystallographic planes in BCC tantalum, another 44 energies and 3564 force training points. 
The defect-free training groups (\emph{Displaced, Volume and Elastic}) therefore contribute a total of 308 energies and 6900 forces compared to the disordered groups (\emph{GSF, Liquid and Surface}) which contributes 54 energies, and 8397 forces.

Once all of these energies and forces are assembled into the training set (Eq. \ref{euclid}, $T$), the atomic configurations need to be translated into the descriptor space (set of bispectrum components, $D$).
However, there are a few hyper-parameters used in the fit that will affect $D$, namely the radial cutoff ($r_{cut}$) and the parameter $J_{max}$ that determines the total number of bispectrum components used.
Multi-element SNAP fits require an optimization of additional hyper-parameters in order to allow for distinct element types to contribute differently to the local density.
However, the present application involves only one element so the only hyper-parameter to search over is the radial cutoff. 
Figure \ref{eerr} shows how the accuracy of the potentials increase with the number of bispectrum components.  
Note that the linear SNAP shows a diminishing improvement in reducing the energy errors as the size of the descriptor space increases.
Because of this trend of diminishing returns, we fix $J_{max}=4$ for the subsequent comparisons.

It is our approach to first search for optimal hyper-parameters that minimize training errors.  An example of this is shown in Figure \ref{hyper} where $r_{cut}$ is varied for both linear and quadratic forms of the SNAP potential in the range $3-12$~\AA. For each $r_{cut}$, $D$ is recalculated and Eq. \ref{euclid} is solved using $\gamma_n=0$ and the identity matrix for ${\bf w}$, giving the unweighted LLS solution for $\boldsymbol\beta$.

For all values of $r_{cut}$, the addition of quadratic terms resulted in lower regression errors in both energies and forces. 
This is expected, since the quadratic SNAP form includes {\small$K(K+1)/2$} adjustable coefficients, in addition to the $K$ coefficients of the linear form, where $K$ is the number of bispectrum components used. 
While the force error is relatively insensitive to $r_{cut}$, there are multiple local minima in the energy errors. The globally optimal values of $r_{cut}$ were found to be 5.672~\AA~and 5.594~\AA~for the linear and quadratic forms, respectively.  Optimized radial cutoffs for each SNAP functional form and value of $J_{max}$ are captured in Supplemental Material Tables S\ref{rcutopt_linear} and S\ref{rcutopt_quadratic}.  In general, we found that the optimal radial cutoff increased with increasing $J_{max}$, but was similar for both linear and quadratic forms. 

Once the hyper-parameters affecting $D$ have been chosen, the next step in the optimization process is focused on the predictive objective functions.
Intuitively, not all of the training points will be equally important for these objective functions.
For example, energies and forces of training configurations corresponding to the generalized stacking fault (GSF) in tantalum will help to accurately reproduce the Peierls stress but will not constrain the bulk modulus, since there is no volume change in these configurations.
How much weight should be given to these training points relative to others is something that can be optimized by adjusting ${\bf w}$ of Eq. \ref{euclid}. 
In order to reduce the complexity of this optimization step, training data is sorted into a limited number of groups, and all rows in $D$ belonging to a group are assigned the same weight.  
In addition, we also assign a different weight to energy and forces rows within a group.
Although the number of free variables is doubled, this is necessary to prevent the LLS fit being dominated by force errors since there are far more rows in $D$ for forces than energies.
\begin{figure*}[t]
\includegraphics{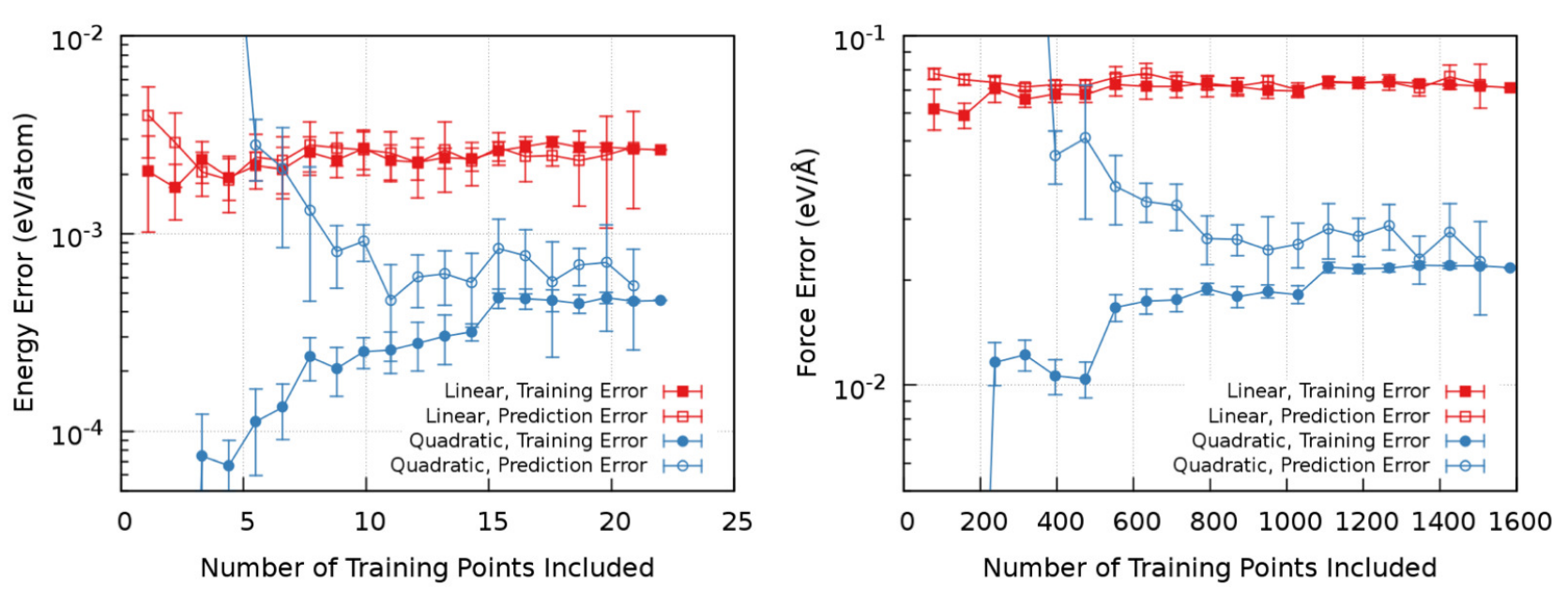} 
\caption{\label{xvalid} 
Comparison of training errors (closed symbols) and prediction errors (open symbols) for the linear (red) and quadratic (blue) SNAP forms.  Different proportions of randomly selected training data are excluded from the fitting process and are used to measure prediction error. Each point in these plots adds an additional 5\% of the total training data, the last point includes all training points and as such has no prediction error to report. Error bars indicate standard deviation across 10 random draws. {\bf Left} Mean absolute energy error for the GSF(110) structures. {\bf Right} Mean absolute force error for the GSF(110) structures. While the quadratic SNAP errors are lower than the linear SNAP form, more training data is needed to fully converge the prediction errors. For training data fractions less than $\sim$ 30\%, prediction errors for the quadratic SNAP form are extremely large because there are fewer training points than fitting coefficients and the resultant potential is strongly overfit.
}
\end{figure*}

It is by no means obvious how best to separate configurations into groups in order to perform this weighted regression.
In Ref. [\onlinecite{Thompson2015}] the training data was grouped according to the phase of tantalum as it was calculated within DFT.
Additional subgroups were formed based on perturbations made to the structure. 
However, there is a more robust way to analyze the similarity of configurations and in turn efficiently assign group weights that aid in optimizing the potential.
In Figure \ref{pca}, the matrix $D$ is projected onto its first two principal components for either the energy (left panel) or force (right panel) rows. 
There is very little distinction between the training groups in terms of the forces. Significant scatter and overlaps are observed between all groups. 
In contrast the first two energy principal components, left panel of Figure \ref{pca}, reveal unique features and similarities of the tantalum training set.
Expectedly, the volumetric expansion and compression of the A15, BCC and FCC phases (diamond symbols) cluster together but liquid structures (pink triangles) are isolated from all other groups. 
Although the authors designated separate groups for either GSF and the free surface configurations, in the energy principal component projection of $D$ these groups do not seem unique. 
This similarity suggests that one can eliminate some free variables during optimization of the group weights. 

Once training groups are defined, we employ a genetic algorithm (GA) within the DAKOTA \cite{Dakota2006} software package to search for optimal group weights.
The quality of a fitted potential is determined by the objective functions specified for the particular application of interest.
This is necessarily open-ended because the same training data can be used to fit many unique SNAP potentials when using a weighted LLS regression model. 
A variety of other approaches could be used to obtain optimal group weights. For example, Bayesian analysis could be used to infer the maximum likelihood values of the group weights.  
We do not expect the overall quality of the SNAP potential to be sensitive to the particular choice of optimization method.
As the GA converges to some set of group weights for a set of objective functions, one can identify which training data was most important to the fit quality. 
Since the calculation of bispectrum components is the computational bottleneck with SNAP, this sequential approach to fitting hyper-parameters first, followed by group weights is significantly more computationally efficient because $D$ does not need to be recalculated for each iterations of the GA optimization loop. 
Being a gradient free method, the GA never fully converges the group weights. 
Typically, we see little variation in fit quality after $\mathcal{O}(10^{4})$ candidate potentials are generated. 
This convergence rate holds for both linear and quadratic forms of SNAP.

\section{\label{prediction}Prediction Beyond Training}
One of the primary concerns with large ANN potentials, which includes quadratic SNAP as a special case, is that the large number of available fitting coefficients may result in a ML-IAP that is overfit to the training data and objectives.  In other words, while the ML-IAP may accurately reproduce data to which it was fit to or evaluated against, it may produce highly spurious results for other points in the descriptor space.
A common way to test for overfitting is to withhold training data and compare the accuracy of the model for these out-of-sample structures (prediction errors) to the accuracy of the model for structures included in training (training errors). 
This is known as cross-validation, and is a staple of robustness testing for ANN models.\cite{Lubbers2017}

In the context of SNAP fitting, cross-validation is achieved simply by excluding a fixed fraction of training data from each training group. This includes both the energies and forces for an excluded configuration.
We are primarily concerned with how much training data is needed to converge prediction errors for excluded data to the same level as the training errors for included data.
To do so, we incrementally increase the fraction of training data included in 5\% intervals and at each fraction ten random selections of included data are generated for each group.
Note, the energy and forces from a single DFT calculation cannot be split across the included/excluded sets.
To exemplify this process, Figure \ref{xvalid} shows training and predictions errors for the GSF(110) group.
Error bars represent the standard deviation among the ten random draws at each fraction included.  
The GSF(110) group is made up of 22 unique DFT calculations each with 24 atoms, totaling 22 energy and 1584 force data points. 
The left panel of Figure \ref{xvalid} represents the energy errors while the right panel captures the force errors observed in this cross-validation. 

Reiterating what was discussed earlier and shown in Figures \ref{eerr} and \ref{hyper}, while the quadratic SNAP form has the same number of bispectrum components, the overall errors are lower than the linear SNAP form.
This is also true when training data is omitted as seen in Figure \ref{xvalid}. The prediction errors in quadratic SNAP are lower than linear SNAP after only 30\% of the total data is used in fitting.
However, the prediction errors do not converge to the regression errors as quickly for quadratic SNAP. 
This points to a need for larger training sets when exercising the expanded set of fitting coefficients available in quadratic SNAP in order to prevent overfit solutions.
Interestingly, this convergence rate of cross-validation errors varies between training groups.
More complex configurations such as GSF, surfaces, and liquids require more training data in order to make accurate predictions for new configurations. 
Simpler groups, such as high symmetry crystal structures, make excellent predictions with very little training.

We see this variation in data requirements as an effect of the chosen descriptor space.
From our experience the bispectrum components easily resolve areas of the PES related to crystalline phases, but require more training in disordered and over/under-coordinated atomic environments. 
Unsurprisingly, these types of training data that we identify as important are also computationally the most costly to acquire from DFT.
This cross-validation, in conjunction with the principal component analysis, could be done prior to group weight optimization and be used as feedback where additional DFT calculations should be made. 
Therefore, the training set construction is done in a way that expands $D$ in a physically meaningful way to improve the robustness of the resultant potential.

\section{\label{sec:level1}Conclusions}
The promise of machine learning for interatomic potentials is enormous. This is due to the fact that once a methodology has been developed and
validated for one particular material or chemical system, it can easily be extended to many more systems, subject only to the availability of appropriate high accuracy training data.  This is in stark contrast to traditional interatomic potentials, where successful development of an IAP for one particular system is not easily extended to others.
For this reason, there is significant activity within the computational physics, chemistry, and materials science communities to not only generate new ML-IAPs for a range of materials and applications, but also to develop new descriptors and representations that are capable of robustly describing atomic potential energy landscapes with an accuracy approaching that of the underlying quantum electronic structure calculations.
However, the common pitfall that plagues many ML-IAPs is their inability to make accurate predictions for configurations that were not explicitly included in the training data.

We have shown that augmenting the SNAP potential to include quadratic terms in the bispectrum components of the neighbor density transforms SNAP from a four-body to a seven-body potential and in turn is able to more accurately capture energies and forces from a diverse training set. 
Interestingly, in the context of Gaussian process prediction of forces in crystalline silicon structures, Glielmo \emph{et al.} recently demonstrated a similar improvement when
replacing a two-body linear kernel with a three-body quadratic kernel \cite{Glielmo2017}
The quadratic SNAP form can be seen as a special case of an ANN with two hidden layers, while the linear SNAP form has only one hidden layer. 
Similarly to ANN potentials, the quadratic SNAP form requires substantially more training data in order to prevent overfitting, as demonstrated
using cross-validation analysis with an extensive set of DFT training data for tantalum structures. 
This large accuracy improvement, an order of magnitude reduction for training error, incurs less than a twofold increase in computational cost.

\begin{acknowledgments}
We would like to thank Art Voter, Danny Perez, Tom Lenosky and Alex Shapeev for fruitful discussions relevant to this manuscript.
This research was supported by the Exascale Computing Project (17-SC-20-SC), a collaborative effort of the U.S. Department of Energy Office of Science and the National Nuclear Security Administration.
Sandia National Laboratories is a multi-mission laboratory managed and operated by National Technology and Engineering Solutions of Sandia, LLC, a wholly owned subsidiary of Honeywell International, Inc., for the U.S. Department of Energy's National Nuclear Security Administration under contract DE-NA0003525.
\end{acknowledgments}

\widetext
\begin{center}
\textbf{Supplemental Material}
\end{center}
\setcounter{equation}{0}
\setcounter{figure}{0}
\setcounter{table}{0}
\setcounter{page}{1}
\makeatletter
\renewcommand{\theequation}{S\arabic{equation}}
\renewcommand{\thefigure}{S\arabic{figure}}
\renewcommand{\bibnumfmt}[1]{[S#1]}
\renewcommand{\citenumfont}[1]{S#1}

\renewcommand{\tablename}{Table S\@gobble}
\begin{table}[h]
\caption{Optimal radial cutoff distances for the linear SNAP form. The corresponding mean absolute energy errors are the same as those plotted in Figure \ref{eerr}.}
\begin{ruledtabular}
\begin{tabular}{p{1.9cm}p{2.0cm}p{1.9cm}p{1.9cm}p{1.9cm}}
$J_{max}$&$r_{cut}$ (\AA)&\hspace{-2em}$\delta E$ (eV/atom)\\
\hline
1&3.940&0.71663\\
2&4.589&0.13793\\
3&5.493&0.04545\\
4&5.672&0.03041\\
5&5.412&0.02446\\
6&5.149&0.02040\\
7&5.506&0.01939\\
\end{tabular}
\end{ruledtabular}
\label{rcutopt_linear}
\end{table}

\begin{table}[h]
\caption{Optimal radial cutoff distances for the quadratic SNAP form. The corresponding mean absolute energy errors are the same as those plotted in Figure \ref{eerr}.}
\begin{ruledtabular}
\begin{tabular}{p{1.9cm}p{2.0cm}p{1.9cm}p{1.9cm}p{1.9cm}}
$J_{max}$&$r_{cut}$ (\AA)&\hspace{-2em}$\delta E$ (eV/atom)\\
\hline
1&3.575&2.04045\\
2&5.861&0.16786\\
3&5.093&0.03229\\
4&5.594&0.00898\\
5&7.039&0.00163\\
\end{tabular}
\end{ruledtabular}
\label{rcutopt_quadratic}
\end{table}

\end{document}